\documentclass[rmp,twocolumn,superscriptaddress,aps,amsmath,amssymb,graphicx]{revtex4-1}

\usepackage{color}
\usepackage{amsmath}    
\usepackage{amssymb} 
\usepackage{amsfonts}   
\usepackage{graphicx}
\usepackage{pgfplots}
\usepackage{bm}

\usepackage{natbib}
\setcitestyle{authoryear,open={(},close={)}}

\definecolor{RawSienna}{cmyk}{0,0.72,1,0.45}
\definecolor{Dandelion}{RGB}{255,212,100}
\definecolor{Emerald}{RGB}{4,99,7}
\definecolor{Grey}{RGB}{50,50,50}
\definecolor{LightGrey}{RGB}{150,150,150}
\definecolor{DarkRed}{RGB}{230,130,130}

\newcommand{\TM}{\color{black}}
\newcommand{\s}{\sigma}	
\newcommand{\bs}{\bm{\s}}

\renewcommand{\>}{\rangle}
\newcommand{\<}{\langle}

\newcommand{\cfeq}{eq.~}

\newcommand{\eg}{\textit{e.g. }}
\newcommand{\ie}{\textit{i.e. }}

\begin{document}

\title{Modeling the correlated activity of neural populations: A review}

\author{Christophe Gardella}
\affiliation{Laboratoire de physique statistique, CNRS, Sorbonne
  Universit\'e, Universit\'e
  Paris-Diderot, and \'Ecole normale sup\'erieure (PSL University),
  75005 Paris}
\affiliation{Institut de la Vision, INSERM, CNRS, and Sorbonne
  Universit\'e, 75012 Paris}
\author{Olivier Marre}
\affiliation{Institut de la Vision, INSERM, CNRS, and Sorbonne
  Universit\'e, 75012 Paris}
\author{Thierry Mora}
\affiliation{Laboratoire de physique statistique, CNRS, Sorbonne
  Universit\'e, Universit\'e
  Paris-Diderot, and \'Ecole normale sup\'erieure (PSL University),
  75005 Paris}

\begin{abstract}
 % abstract
The principles of neural encoding and computations are inherently collective and usually involve large populations of interacting neurons with highly correlated activities. While theories of neural function have long recognized the importance of collective effects in populations of neurons, only in the past two decades has it become possible to record from many cells simulatenously using advanced experimental techniques with single-spike resolution, and to relate these correlations to function and behaviour. This review focuses on the modeling and inference approaches that have been recently developed to describe the correlated spiking activity of populations of neurons. We cover a variety of models describing correlations between pairs of neurons as well as between larger groups, synchronous or delayed in time, with or without the explicit influence of the stimulus, and including or not latent variables. We discuss the advantages and drawbacks or each method, as well as the computational challenges related to their application to recordings of ever larger populations.

\end{abstract}

\maketitle

\tableofcontents{}

 % 0_intro
\section{Introduction}

One of the main goals of neuroscience is to understand how the collective activity of neural networks is influenced by sensory inputs and underlies animal behaviour.
Most studies in the past century have studied how single neurons represent information and perform computations. Over the last two decades, technological progress has allowed scientists to record increasingly larger populations of neurons, up to complete organisms \citep{Ahrens2013whole,Prevedel2014,Wolf2015whole,Dupre2017}. These recordings have revealed a complex picture where neural networks cannot be solely understood from their individual components.

Anatomically, neurons form highly distributed networks. Cortical pyramidal neurons for example can connect to and receive input from up to thousands of neurons \citep{Shepherd2004}. This connectivity results in complex dynamics, where some neural ensembles tend to respond synchronously. Even in the absence of sensory input, spontaneous activity in sensory areas can take the form of large synchronous oscillations.
Information is now thought to be represented not just by the firing rate of specific cells, but by attractors of neural response trajectories \citep{Hopfield1982}, or even by trajectories themselves \citep{Mazor2005}.
Although neural populations are largely redundant \citep{Warland1997,Puchalla2005}, some information is synergistic, and cannot be recovered by recording single cells \citep{Schneidman2011,Franke2016,Zylberberg2016direction}.

Whereas isolated neurons are reliable components \citep{Mainen1995,Nowak1997influence}, neural network have intrinsic variability \citep{Deweese2004,Stringer2016inhibitory}.
Part of this variability might result from noise building up along successive computations \citep{Faisal2008noise}.
Variability might also be due to internal network dynamics and top-down modulation \citep{Arieli1996,vanVreeswijk1996,Vogels2005neural,Nienborg2012decision}.
This correlated variability might be more than just noise.
Correlated fluctuations of a neuron's input, even independent of the stimulus, can increase the reliability of its response to weak stimuli \citep{Shu2003barrages}, an effect called stochastic resonance \citep{Longtin1991time}. 
Correlation between excitatory and inhibitory inputs can also balance neural activity and increase temporal precision \citep{Wehr2003balanced,Cafaro2010,Baudot2013animation}.
Correlated activity also facilitates signal transmission across neural networks, even for uncorrelated stimuli \citep{Reyes2003}.
Some works even suggest that stimulus-dependent noise correlations could improve the precision of stimulus encoding \citep{Franke2016,Zylberberg2016direction}. 
Thus, in order to understand how neurons enco	de and process information, it is important to characterize their correlated activity.

Here we review multiple approaches that have been proposed to build explicit probabilistic models of the correlated response of populations of neurons.
We focus on models of spike trains rather than of firing rates, as multiple works have shown that spike patterns contain more than just their rate \citep{Wehr1996,Berry1997,Reich1997response,Reinagel2000temporal,Panzeri2001,Nemenman2008,Kayser2009,Baudot2013animation}. Spike trains are also a more faithful description of neurons activity, as organisms sense and act in single trials.
Spike train models have been used to study multiple features of the neural code in multiple circuits, such as noise correlations in the auditory cortex \citep{Lyamzin2015}, neural dynamics in primary motor, parietal and ventral premotor cortices \citep{Truccolo2010}, error-correcting stimulus encoding and interactions between neurons in the retina \citep{Schwartz2012b,Schneidman2006,Ganmor2011architecture} and in the primary visual cortex \citep{Ohiorhenuan2010},.

We focus primarily on models whose parameters can be inferred directly from neural recordings.
If a model is too simple, it might not capture some structure in data. But if it is too complex, it will be hard to infer and will require a lot of experimental data.
In the models we present, neural responses are influenced by three sources: the spiking history of the recorded population, the stimulus, and latent variables representing unobserved sources such as neighboring neurons that were not recorded. The model chosen for a study depends on the pursued goal.
If the aim is to decode a stimulus from the neural activity, the model should take this stimulus into account.
The model also depends on the recorded area: strong modulations by internal dynamics, modeled for example by dynamical latent variables, might be more adapted to higher level areas such as the prefrontal cortex than to lower level ones such as the cochlea.

The models we review describe correlations in a population of $N$ neurons. In order to have a unified framework for all models presented here, we represent spike trains as binned responses. The time is divided in bins of equal length, and the response of neuron $i$ in time bin $t$ is represented by the variable $\s_{it}$.
Depending on the model, $\s_{it}$ can either be binary (1 if there is at least one spike in the bin, 0 otherwise) or an integer corresponding to the number of spikes in the bin. We refer to the former representation as binary responses, and to the latter as spike count responses.
The population responses in time bin $t$ is then represented by a vector of dimension $N$, $\bm{\s}_t=(\s_{1t},\ldots,\s_{Nt})$, where the bold font is used for vectors and matrices.
Depending on the goal of the study and on the neural population under consideration, the size of time bins can be very different, typically from 1 ms \citep{Pillow2008,Truccolo2010} to longer than a second \citep{Franke2016}. 
 Some of these models, such as the generalized linear model, can also be expressed in continuous time, corresponding to a vanishing bin size in our adopted framework.

 % 1_singlebin
\section{Collective synchronous activity}
Neurons in a population are correlated both in time and across neurons.
We begin by presenting models of neural responses in a single time bin, which focus on synchronous activity. Synchronous activity has received a lot of attention \citep{GraySinger95}. Several models aim at telling which neurons fire synchronously \citep{Grun2002,Shlens2009,Cocco2009a,Stevenson2009}.

It should be noted that these models do not assume that temporal correlations are absent, but rather choose to focus of synchronous correlations. However, in some cases where correlations between neurons concentrate on short times that are comparable to the choice of the time bin, these models do in fact provide complete and accurate description of the temporal activity.

The number of possible responses grows exponentially with the population size. Even for populations of a few tens of neurons, the distribution of responses is thus impossible to probe experimentally. 
Multiple models have been proposed, with different degrees of precision. We start with coarse models, with low numbers of parameters, and progressively increase their complexity. Unless explicitly mentioned, all models of synchronous activity are models of binary responses.

\subsection{Stimulus-independent models of activity}
\label{s:model_simple}

Stimulus-independent models represent the correlated activity of a neural population with no information about the stimulus. This does not mean that correlations between neurons are not due to stimulus, but only that the model does not use that information, just as the brain does not since it has no direct access to it. This can be convenient if no model is available to relate neural responses to stimuli, or if not enough experimental data is available to learn such a model, or if one wants to think about how the nervous system can build representations of neural activity.

\subsubsection{Coarse models}

Before presenting specific models, we begin by presenting the framework of Maximum Entropy (ME) models. Many of the models presented below are derived from this framework.

ME models are a powerful tool for modeling distributions from which only limited samples can be observed \citep{Jaynes1957}. 
Here, we use the ME principle to model the distribution of neural responses $P(\bm{\s})$.
The core idea of ME is to search for the most random distribution that reproduces some chosen descriptive statistics. The model is designed so that these statistics have the same value in the model and in data.
One first picks several observables of interest, $\mathcal{O}_n (\bm{\s})$, and computes their empirical mean:
\begin{equation}
\< \mathcal{O}_n(\bm{\s})\>_\text{emp} = \dfrac{1}{T} \sum_{1}^{T}  \mathcal{O}_n( \bm{\s}_t )
\end{equation} 
where $T$ is the number of samples. For example, the mean of $\mathcal{O}_n(\bm{\s}) = \s_i$ is the mean response of neuron $i$. 
Then one searches for the model of maximum entropy, $S(P) = - \sum_{\bm{\s}} P(\bm{\s}) \log P(\bm{\s})$, amongst all models that reproduce these statistics. 

Given the constraints, the choice of maximum entropy can be interpreted as making as few assumptions as possible to avoid introducing biases.
The ME model minimizes the worst-case cost of estimation errors, as measured by the log-loss \citep{Grunwald2004}. For the special case of pairwise models applied to retinal data (see below), it was shown that the ME model outperformed other models picked at random within the same constraints, especially for large networks \citet{Ferrari2017random}.

Maximum entropy distributions can sometimes be viewed as resulting from ``shuffling'' experiments, where the patterns of activity are scrambled across times and neurons while keeping the constrained statistics constant \citet{Socolich2005,Okun2015}. The equivalence of this method with maximum entropy has been proven in the special case of pairwise correlations \citet{Bialek2008a}, and is likely to be general. While it provides a practical way of generating new activity patterns, unlike ME models this technique cannot evaluate the probability of given patterns, and is unable to extract phenomenological parameters such as coupling constants.

By maximizing the entropy using the technique of Lagrange multipliers to enforce the contraints, one can show that the ME model has the following form:
\begin{equation}
P(\bm{\s}) = \dfrac{1}{Z} \exp \left( \sum_n \theta_n \mathcal{O}_n(\bm{\s}) \right)
\label{eq:ME}
\end{equation}
where $\theta_n$ are the Lagrange multipliers. $Z$ is a normalization constant called partition function, which can be hard to infer, but is only needed for specific applications.  
For conciseness, 
in the following we will often omit $Z$ and use instead a proportionality sign $\propto$.
Solving for the constraints is equivalent to
maximizing the likelihood of the observed data, which is a convex problem. By choosing different observables and comparing the corresponding ME model to empirical responses, one can quantify interactions at different orders \citep{Bohte2000,Schneidman2003}, and estimate the mutual information between stimulus and responses \citep{Schneidman2003,Montemurro2007}.

Importantly, although ME model have a common form (\cfeq\ref{eq:ME}), they can be very different in practice, depending on the constrained observables. As we will see, some models can be solved exactly from data, while others require advanced computations to approximate them.

Perhaps the simplest description of the collective population activity is the total number of spiking neurons:  $K(\bs) = \sum_{i=1}^N \s_i$.  In the literature, $K$ is sometimes termed population rate, although it is more of a population count.
\citet{Tkacik2013a} proposed to model the distribution of responses with the maximum entropy model reproducing the distribution of the spike count, $P(K)$. This model takes the form:
\begin{equation}\label{pk}
P(\bm{\s}) \propto   \exp[ g(K(\bs))]
\end{equation}
where the function $g(K)=\ln P(K)-\ln \binom{N}{K}$ is learned from data. This model is very easy to learn, even for large populations, using only the histogram of the population count $P(K)$. It is equivalent to the \textbf{homogeneous neural pool model} \citep{Amari2003}, where all neurons are considered identical. It can be shown that it is the only model reproducing the spike count while being invariant to permutations of neurons.
\citet{Montani2009} used this model to study higher order correlations (\ie correlations involving more than two neurons) in somatosensory cortex. They showed that interactions between pairs and triplets of cells were sufficient to account for the distribution of the population count $K$ in a population of 24 cells. It remains unclear how this model could also account for the fine structure of correlations between neurons, as it assumes that all neurons are equivalent.

Another simple ME model is one where the mean activity of each neuron is constrained, $\<\s_i\>$. These constraints lead to a \textbf{model of independent neurons}:
\begin{equation}	
 P(\bm{\s}) \propto \exp\left( \sum_i h_i \s_i\right)=\exp( {\bm h}^\intercal \bs) 
\end{equation}
where the coefficients $h_i=\ln[\<\s_i\>/(1-\<\s_i\>)]$ control the mean spiking of each neuron. These coefficients are sometimes called fields, by analogy with magnetic fields in physics. This model is very convenient to learn, to generate data or to compute responses likelihood. On the other hand, it is quite limited, as it cannot account for any correlation. This model is often considered as a default model, setting the baseline for the comparison of other models \citep{Macke2011,Koster2014}.

\subsubsection{Models of population coupling}
These coarse models can be combined to define a more general ME model that can still be solved with relative computational ease: the \textbf{population coupling model}, which reproduces the joint probability between each neuron $i$ and the population count, $P(\s_i, K)$ \citep{Gardella2016}. 
This model is motivated by recent work suggesting that some neurons are sensitive to the global population activity rather than to the detailed activity of individual neurons \citep{Okun2015}. This model was also proposed elsewhere, in a slightly different form, where it was termed population tracking model \citep{Cian2017}.

The model takes the form:
\begin{equation}
 P(\bm{\s}) \propto \exp[\bm{h}_{K(\bs)}^\intercal \bm{\s}]
\end{equation}
where the field vector $\bm{h}_{K}$ depends on the value of the population count $K$. The population activity influences the activity of each cell. 
Conveniently, this model is tractable in the sense that all predictions, as well as the partition function, can be obtained with arbitrary precision using only polynomial operations that scale with $N^3$. 

The model can be used to measure how much the population activity shapes the precise structure of pairwise correlations, $\<\s_i\s_j\>-\<\s_i\>\<\s_j\>$. In the salamander retina \citep{Gardella2016}, the population activity explained 50\% of pairwise correlations, which was similar to what was reported in mouse V1 \citep{Okun2015}.
Recent application of the population coupling model to the human and monkey cortices showed that the model explained the collective activity well during sleep, but that detailed interactions between specific pairs of neurons mattered during wakefulness \citep{Nghiem2018}. SImilarly, in mouse V1, population rate predicted better pairwise correlations during synchronized states, and very little during desynchronize states \citep{Okun2015}. 
Thus, while modeling interactions with the global activity drastically simplifies the structure and inference of the distribution of activities, but it provides only a coarse description of responses which may be inappropriate or incomplete depending on the neural context.

A simpler model that only reproduces the distribution of the population count and its coupling with each neuron, $\< K(\bs) \s_i \>$, has also been proposed \citep{Gardella2016}. Interestingly, for a population of 160 neurons in the salamander retina this simpler model predicted pairwise correlations almost as accurately as the complete coupling model, with 50 times less parameters. 

\subsubsection{Models of pairwise correlations}

Models based on the total population count and individual activities are easy to infer, but they can only provide a limited description of responses. A finer description of the structure of responses is the correlation between pairs of neurons. Several models have been proposed to describe these pairwise correlations.

The maximum entropy model reproducing the firing rates of all cells, $\< \s_i \>$, and the correlation between all pairs of neuron $\< \s_i \s_j \>$ has been used extensively to study correlations in the retina or in multiple areas of the brain \citep{Schneidman2006,Tang2008,Koster2014,Meshulam2017}. This model is often called the \textbf{Ising model} and is equivalent to the ``Boltzmann Machine'' introduced by \citet{Hinton1984}. It takes the form:
\begin{align}
P(\bm{\s}) &\propto \exp\!\left(\sum_i h_i \s_i + \sum_{j\neq i} J_{ij} \s_i \s_j \right) \\ 
&\propto\exp\left(\bm{h}^\intercal \bs + \bm{\s}^\intercal \bm{J \s} \right)
\label{eq:Ising2}
\end{align}
where \cfeq\ref{eq:Ising2} is written in matrix form. $\bm{h}$ is a line vector of field coefficients, and $\bm{J}$ is a symmetric matrix of interaction couplings. The Ising model is very similar in structure to the Hopfield model \citep{Hopfield1982}, a classical model of memory storage.

The performance of the Ising model at representing the distribution of responses is little affected by pruning weak edges \citep{Ganmor2011architecture} (\ie setting small $J_{ij}$ to 0), or by limiting interactions to neighbors only \citep{Shlens2006,Shlens2009}.
The Ising model is less accurate when responses have strong correlations, \eg for neighboring cortical cells \citep{Ohiorhenuan2010} or at short time scales where refractory periods have strong influence on neuron's dynamics \citep{Bohte2000}.

The Ising model works best on relatively small population, for which its number of parameters remains reasonable. It can be used as a building block for coarse-grained models that couple uniform sub-populations of neurons. In the \textbf{hierarchical model} \citep{Santos2010}, neurons are partitioned into $n$ pools $I_1, ..., I_n$.
Pools are formed such that neurons in each pool are as homogeneous as possible, and are each modeled by an Ising model. Then, interactions between pools only depend on their population counts $K_{1}, ..., K_{Q}$, with $K_a = \sum_{i \in I_a} \s_i$:
\begin{equation}
P(\bm{\s}) = Q( K_1, ..., K_Q )  \prod_I P(\bm{\s}_I)
\end{equation}
where $\bm{\s}_I$ is the sub-vector of neurons in pool $I$. The function $Q$ of interaction is given by a ME model corresponding to various choices of constrained observables, \eg pairwise joint distribution between the spike counts of two pools.
\citet{Santos2010} found that for cortical cultures, the peaks of local field potentials (LFP) activity were better described by the hierarchical model than by the Ising model.
Such hierarchical models could also prove useful to model the interactions between neurons from multiple areas.

The Ising model is a powerful theoretical concept, and it can reproduce any pairwise correlations exactly, but it is notoriously hard to infer for populations larger than a few tens of neurons, even with advanced computational techniques  \citep{Cocco2009a,Ferrari2016learning}. 
Generating data with this model is also computationally demanding, as it requires complex techniques such as Monte Carlo methods. However, computing 
responses likelihood is easy, up to the normalization constant. This normalization constant is hard to estimate, but it is not always needed, for example if one only wants to compare the likelihoods of different responses.

An alternative to the Ising model for modeling correlations is the \textbf{cascaded logistic model}. This model is not equivalent  to the Ising model but it is close in some cases with sparse connectivity \citep{Park2013}. This model is build by conditioning the activity of neurons on previous ones using logistic models: 
\begin{equation}
P(\bm{\s}) = P(\s_1) \prod_{i>1} P(\s_i|\s_1 ... \s_{i-1})
\end{equation}
with 
\begin{equation}
P(\s_i=1|\s_1...\s_{i-1}) = f\left( h_i + \sum_{j<i} w_{ij} \s_j \right)
\label{eq:CLM}
\end{equation}
where $ f(x) = 1/(1+e^{-x})$ is the logistic function. The interaction weights $w_{ij}$ are learned from data using maximum likelihood estimation. This model is sensitive to the order of neurons, but algorithms have been developed to find orders that make this model close to the Ising model \citep{Park2013}. 
Although inferring this model is easier than the Ising model, it is still not trivial. Unlike the Ising model, it cannot reproduce all correlations exactly, and may even get the mean activities very wrong. The cascaded logistic model is very  convenient to generate data and to calculate the likelihood of responses.

Other methods have been proposed to generate responses with fixed pairwise correlations. Both models we present here are \textbf{common input models}, where responses are influenced by a latent input process.  Although these models are not guaranteed to exist for any given covariance matrix, such methods are relatively simple theoretically and convenient to generate data. 

A first method is the \textbf{spike train mixture}. Several similar techniques have been proposed, which have latter been unified and generalized by \citet{Brette2009}.
The mixture method was initially introduced for spike trains without response binning. We present it here in the context of binned spike count responses for consistency. 
One starts by generating $M$ latent Poisson processes, called spike train sources, with rates $\nu_1, ..., \nu_M$. Then, each spike of source $j$ is inserted in the response of neuron $i$ with probability $w_{ij}$. The response of neuron $i$ is then a Poisson process with rate:
\begin{equation}
\lambda_i = \sum_j w_{ij} \nu_j
\end{equation}
and with correlation between neurons $i\neq i'$:
\begin{equation}
\< \s_i \s_{i'} \> = \sum_j w_{ij} w_{i'j} \nu_j.
\end{equation}
\citet{Brette2009} presented methods to find appropriate rates $\nu_1, ..., \nu_M$ and mixing matrix $\bm{w}$ given target firing rates and covariances. 
Unfortunately, such a solution may be degenerate, and does not always exist for all target correlations; in particular the model cannot produce negative correlations.
This model is not equivalent to the Ising model and it introduces additional effective higher-order interactions. The likelihood of responses under the model is not straightforward to compute.

The \textbf{dichotomized Gaussian model} \citep{Amari2003}, another popular common input model for pairwise correlations, is easy to use in practice. Neurons are driven by a Gaussian latent variable $\bm{z} \sim \mathcal{N}(\bm{\mu, C})$ of dimension $N$. 
Neuron $i$ has a binary response: it spikes if $z_i$ is positive, else it is silent.
A more general model has also been proposed allowing for non-binary spike counts with arbitrary distributions \citep{Macke2009}. \citet{Macke2009} presented efficient ways to find parameters $\bm{\mu}$ and $\bm{C}$ agreeing with the data, although such a solution does not always exist, for example in some cases with strongly negative correlations.

Like the spike train mixture model, this model implies higher-order interactions and is distinct from the Ising model, although the importance of this difference is still debated. \citet{Macke2009} showed that for many parameters, the entropy of the dichotomized Gaussian model is close to the entropy of the maximum entropy model, compared to the spike train mixture model.
On the other hand, {\TM \citet{Yu2011}} found that when modeling negative deflations in local field potential (LFP) in the macaque premotor cortex, the dichotomized Gaussian and Ising models made different predictions for response statistics, and the former made predictions closer to data than the latter.
Recently \citet{Lyamzin2015} used a dichotomized Gaussian model to explain the dependence between stimulus and noise correlations observed \textit{in vitro} in mice and \textit{in vivo} in gerbil auditory cortex.

All four models of pairwise correlations ---Ising, cascaded logistic, spike train  mixture, and dichotomized Gaussian---
have about the same number of parameters, but very different computational costs for inference.
The most random model, \ie the Ising model, is surprisingly the hardest to train. When the goal is to reproduce mean activities and pairwise correlations exactly, with the guarantee that no higher-order interactions are present, the Ising model should be preferred. The Ising model can also be used to study the structure of the probability distribution, such as its basins of attraction \cite{Tkacik2014b,Watanabe2014a,Loback2018}, or thermodynamic properties (\eg closeness to a critical point, see \citet{Tkacik2015}). On the other hand, if one simply needs to generate responses with a model able to reproduce correlations in most cases, the dichotomized model may be preferred.

\subsubsection{Adding population coupling to pairwise models}
It was recently observed that for large populations of over 100 neurons in the salamander retina, the Ising model does not predict interactions between neurons  higher than second order \citep{Ganmor2011sparse,Tkacik2014b}. 
For example, it fails to estimate the distribution of the population count, over-estimating by more than an order of magnitude the probability of high values of the population count $K$. 
Several works have attempted to correct this bias. 

The most straightforward strategy to correct the Ising model is to add the distribution of the population count to the set of observables constrained by the ME model.
The resulting \textbf{$K$-pairwise model} is the maximum entropy model reproducing the mean spiking activities, pairwise correlations, population count histogram $P(K)$ \citep{Tkacik2014b}:
\begin{align}\label{pkpairwise}
 P(\bm{\s}) \propto \exp\left[\bm{h}^\intercal \bs + \bm{\s}^\intercal \bm{J} \bm{\s} + g(K(\bs)) \right]
\end{align}
where the function $g$ is learned jointly with $\bm{h}$ and $\bm{J}$ from data. Learning this model on the activity of the salamander retina improved the prediction of triplet correlations over the Ising model.

Another model generalizing the Ising model is the \textbf{semiparametric energy-based probabilistic model} where the exponential in \cfeq\ref{eq:Ising2} is replaced by an arbitrary increasing function $f$ learned from data along with the fields $\bm{h}$ and couplings $\bm{J}$ \citep{Humplik2016}:
\begin{equation}
P(\bm{\s}) \propto f  \left( \bm{h}^\intercal \bs + \bm{\s}^\intercal \bm{J} \bm{\s} \right).
\label{eq:semipar}
\end{equation}
A non-exponential $f$ can be useful to correct the errors of the Ising model, for example by decreasing the probability of rare events such as high population counts. Such a correction can be applied to any maximum entropy model, and is not specific to the Ising model. 
\citet{Humplik2016} showed that the semiparametric model indeed described salamander retinal ganglion cell responses better than the Ising model, as measured by data likelihood. This difference increased with the population size. This was achieved even though the semi-parametric model nonlinearity $f$ was parametrized with a low number of parameters.

Unlike the $K$-pairwise model, the semi-parametric model is not guaranteed to reproduce the mean activities and pairwise correlations, and its parameters are less easily interpretable.
It should also be noted that the good performance of both models in the retina is not guaranteed to hold in cortical networks, where correlations are stronger.

\subsubsection{Higher-order interactions}
\label{sss:higherorder}
Higher-order correlations in neural populations can be hard to model, yet they can have a strong influence on the structure of responses, \eg for neighboring cortical cells {\TM \citep{Ohiorhenuan2010}}.
The models presented so far have relatively simple interactions, either pairwise interactions, or interactions pooling large numbers of neuron together, regardless of their identity. This restriction is set by the amount of experimental data, which in turn limits the number of parameters. And as we will see now, modeling higher order correlations can require large numbers of parameters.

\citet{Martignon1995} proposed to measure interactions between neurons by writing models in the following form:
\begin{equation}
 P(\bm{\s})\! \propto \! \exp\!\bigg( \! \! \sum_i \theta_i \s_i + \sum_{i_1i_2} \theta_{i_1i_2} \s_{i_1} \s_{i_2} + ...  + \theta_{1...N}\s_1 ...\s_N \! \!\bigg)
 \label{eq:theta}
\end{equation}
This form, called full log-linear model or $\theta$-coordinates system, is general in the case of binary variables: any distribution can be written in this form. The interaction of order $m$ between neurons $i_1, ..., i_m$ is set by the coefficient $\theta_{i_1...i_m}$, with a value 0 corresponding to no interaction. The ME model reproducing correlations up to maximum order $m$ has all coefficients for higher order than $m$ set to 0, defining a hierarchy of models as $m$ increases \citep{Schneidman2003}.
In this hierarchy, coefficients of order $m$ quantify interaction between neurons that cannot be explained by lower order interactions. For example, the Ising model only reproduces pairwise correlations, so it has no coefficient of order higher than 2. Writing a model in this form can be convenient to estimate interactions between small groups of neurons \citep{Yu2011}.

The situation is more complex for large populations. As the $\theta$-form is just a re-writing of response distributions, it has the same number of degrees of freedom, $2^N-1$,  and it is often impossible to estimate the coefficients empirically. 
Models have thus been proposed to lower the number of interactions to estimate. A first idea could be to make coefficients  $\theta_{i_1...i_m}$ only depend on the number of neurons, $m$, but not on their identity. This model is invariant to permutations of neurons, and is equivalent to the homogeneous pool model, \cfeq\ref{pk}. If permutation invariance is only required for terms of order 3 or more, then one obtains the $K$-pairwise model of \cfeq\ref{pkpairwise}.

Another way to limit the number of parameters is to include interaction terms without assuming their order \textit{a priori}. In the \textbf{reliable interaction model}  {\TM \citep{Ganmor2011sparse}}, all coefficients are set to 0, except the coefficients necessary to fit the probability of most frequent responses (\ie the responses $\bs$ occurring at least $n_\text{RI}$ times in recordings). The coefficients are then determined by a system of linear equations matching the model and data values of $\ln P(\bs)$ for these responses. For example, denoting by $0$ the state with no spiking neuron, the log-probability that only neuron $i$ spikes is $\theta_i + \log P(0)$; the log-probability that only neurons $i$ and $j$ spike is $\theta_i + \theta_j + \theta_{ij} + \log P(0)$, \textit{etc}.

This model is convenient because it makes no \textit{a priori} assumption about the maximum order of interactions, and does not require to learn all low order interactions. For a population of approximately 100 neurons in the salamander retina, the reliable interaction model was a more accurate description of frequent responses than the Ising model, with 10 times less parameters \citep{Ganmor2011sparse}. 

This model has several drawbacks. The number of interactions involved in the $n_\text{RI}$ most frequent responses is likely to grow exponentially with the population size, which would limit the number of responses used for learning. 
The coefficients not involved in frequent responses are assumed to be 0, but this is not constrained by data, which could lead to dramatically over-estimating the probability of unobserved responses, because strongly negative coefficients would be needed to lower the probability of such responses. 
The model is also not constrained to be normalized.
As a result, the model cannot be used to estimate data likelihood, nor to generate patterns of activity, and is limited to a descriptive usage.

\subsubsection{Restricted Boltzmann machines}
As we have seen, it is hard to model explicitly general higher-order correlations between neurons. An alternative way to model complex interactions is to introduce latent (\ie not observed) variables (called hidden units when they are discrete) each interacting with several neurons. Neurons interacting with a common latent variable are effectively correlated through it. Latent variables can account for correlations of different orders. Although latent variables can sometimes be interpreted as originating from an unobserved input, such as common stimuli or unrecorded cells, they do not necessarily correspond to an existing entity. As we will see, latent variable can drastically simplify models.

The \textbf{restricted Boltzmann machine} (RBM) is a simple model of binary neurons interacting with binary hidden units \citep{Smolensky1986}.
Neurons have pairwise interactions with hidden units. 
While the Ising model is also a Boltzmann machine, this model is said to be \textit{restricted}, as not all pairwise interactions are allowed: there is no direct interaction between two neurons or between two hidden units.  The joint probability between neurons and hidden units is thus:
\begin{align}
 P(\bm{\s, z}) &\propto \exp\left( \bm{a}^\intercal \bm{\s} + \bm{b}^\intercal\bm{z} + \bm{z}^\intercal \bm{w} \bm{\s} \right)
\label{eq:RBM}
\end{align}
where $\bm{z}$ is the column vector of binary hidden units of size $M$, the number of hidden units. Fields $\bm{a}$ and $\bm{b}$ are vectors controlling the activity of neurons and hidden units, and $\bm{w}$ is a matrix of interaction coupllings of size $M \times N$.
$\bm{a}$, $\bm{b}$ and $\bm{w}$ are learned from data by likelihood maximization, which can be done relatively easily using algorithms such as contrastive divergence \citep{Hinton2002} or persistent contrastive divergence \citep{Tieleman2008}.
The probability of responses is obtained by marginalizing over latent variables, which can be computed analytically, up to a normalization constant \citep{Fischer2012}.

The number of hidden units is left to choice, and can be larger than the number of neurons. A compromise must usually be made: the more hidden units, the more complex distributions can be represented, and the more data is needed for inference. Any binary distribution can be approximated with arbitrary precision by an RBM, as measured by the Kullback-Leibler divergence \citep{LeRoux2008}, illustrating the fact that arbitrary complex correlations between neurons can be described by simple interactions with latent variables. 
A very large number of hidden units (and parameters) might be required to achieve arbitrary precision. Because of the risk of overfitting, the RBM is not guaranteed to be an accurate model for experimental recordings of limited duration.
Also, the RBM does not allow for an easy interpretation in terms of interactions between neurons, as the $\theta$-form does (\cfeq\ref{eq:theta}).
 
The restricted nature of the RBM is convenient for computations: given a state of latent variables, neurons are independent, and conversely. 
This make the RBM very convenient to learn and to generate data. The RBM has been shown to be a more accurate model than the Ising model for responses in the cat visual cortex {\TM \citep{Koster2014}}. \citet{Humplik2016} also found that in the salamander retina, the RBM gave a better description of responses compared to the \textit{K}-pairwise and semi-parametric energy-based models, in terms of response likelihood. Similar observations were reported in the rat retina \citep{Gardella2018}. This difference was accentuated for larger population sizes.

The RBM can be enhanced by also allowing interactions between neurons to define a \textbf{semi-restricted Boltzmann machine} \citep{Koster2014}. For populations in the cat primary visual cortex of up to 36 cells, little difference was found between the RBM and the semi-RBM. In this case, the RBM is not only a better model of neural responses than the Ising model, but it is also as efficient as the Ising model at capturing pairwise correlations. 
As the size of the population considered was rather limited, it would be interesting to test if such results also hold for larger populations.

The RBM is thus a very promising model, which outperforms almost all models of synchronous population activity presented so far. Its main drawback is the lack of interpretability of its parameters. It would be interesting to ask whether the performance of the RBM extends to modeling the activity of large populations of cortical neurons in behaving animals.

 % 2_stim

\subsection{Population response to a stimulus}
A significant part of correlations between sensory neurons can be due to the stimulus. Different neurons may respond to common features of the stimulus, and be correlated through this shared influence. Stimulus features may themselves be correlated with other, inducing more correlations downstream.
Including the influence of the stimulus on neural responses should provide a better description of correlated responses.
On the other hand, even for a given stimulus, correlations between the noisy responses may remain. Taking those noise correlations into account can significantly improve the description of responses to stimuli, and thus improve decoding precision \citep{Franke2016,Zylberberg2016direction}.
An extensive part of the neuroscience literature has focused on modeling the influence of a stimulus $\bm{s}$ on responses, $P(\bm{\s}|\bm{s})$. A complete review of such models is beyond the scope of this review. Here, we focus on models of response where correlations cannot be explained by the stimulus alone, or in other words where noise correlations are present.

The framework presented here can describe both correlations with a stimulus or with behaviors \citep{Lawhern2010}. Behaviors can correlate both with motor neurons controlling them, but are also known to modulate sensory neurons \citep{Nienborg2012decision}.
In general, the models presented here correlate neural responses to an external variable $\bm{s}$, without assuming the direction of causality.
We shall refer to $\bm{s}$ as ``stimulus'' for simplicity, but it can also be understood as behavior, such as arm movements.

\subsubsection{Complex noise models}
\label{sss:complex_noise} 
A natural strategy is to model the probability of responses to each stimulus $\bm{s}$, $P(\bm{\s} | \bm{s})$, using any model previously described. This method can model complex noise correlations between neurons. If the set of stimuli is discrete, then the amount of parameters to estimate would grow linearly with the number of stimuli. In order to limit the amount of parameters to estimate, one can impose constraints on the parameters, \eg that some parameters are constant  across stimuli.
In the \textbf{stimulus-dependent Ising model}, each distribution $P(\bm{\s} | \bm{s})$ is modeled by an Ising model (\cfeq\ref{eq:Ising2}). {\TM \citet{Schaub2012}} applied this model to a population of orientation selective neurons, and showed how decoding could be achieved in this framework, using simulations.  But this direct approach used 10000 responses per stimulus, which is challenging experimentally.

Another possible approximation is to consider that in \cfeq\ref{eq:Ising2} fields $\bm{h}$ depend on the stimulus, but not the couplings $\bm{J}$. A challenge is then to model how fields $\bm{h}$ depend on stimulus $\bm{s}$. \citet{Granot-Atedgi2013} modeled it with a Linear Nonlinear model:
\begin{equation}
h_i(s) = f_i( \bm{k}_i^\intercal \bm{s} )
\end{equation}
where $\bm{k}_i$ is a linear filter spanning the stimulus space, and $f_i$ is a nonlinear function.
The success of this strategy was made possible by the simplicity of the stimulus and neurons used: retinal ganglion cells were stimulated by a uniform time-varying light intensity.
But often, there is no simple model to predict how even single neurons respond to a stimulus, \eg natural movies \citep{Gollisch2010,McIntosh2016}.

A helpful trick is to use \textit{time-}dependent models instead of \textit{stimulus-}dependent ones \citep{Tkacik2010,Granot-Atedgi2013,Koster2014,Ganmor2015}. One records multiple responses to repetitions of a stimulus, and computes the mean response for each time bin, $\<\s_{it}\>$, also called peristimulus time histogram (PSTH). Then one studies the ME model that reproduces each $\<\s_{it}\>$, as well as correlations between neurons, $\sum_{t} \<s_{it}s_{jt}\>$:
\begin{align}
P(\bm{\s}_t) \propto \exp\left( \bm{h}_t^\intercal \bm{\s} + \bm{\s}^\intercal {\bm J} \bm{\s} \right)
\end{align}
where fields $\bm{h}_t$ are different for each time bin, and are learned from data along with $\bm{J}$. 
This formulation is convenient, as it makes it possible to study interactions between neurons for any stimulus, even if no model is available to predict responses to stimuli in general. 
Of course, such a model is limited as it requires stimulus repetitions and it cannot make predictions for responses to new stimuli. Nevertheless, a convenient strategy to generalize this model is to first learn time-dependent fields $\bm{h}_t$ and couplings $\bm J$, and combine it with a single-neural model of encoding predicting $\<\s_{it}\>$ to calculate a stimulus-dependent $h_{it}=h_i(\bm{s})$. This avoids optimizing $\bm{J}$ and $\bm{h}(\bm s)$ at the same time, which is computationally challenging and may also wrongly attribute to noise correlations some signal correlations that are missed by the single-neuron model. \citet{Ferrari2018} applied this strategy to populations of retinal ganglion cells responding to moving bars, and showed that the inferred couplings $\bm{J}$ generalized very well to other stimulus ensembles.

\citet{Koster2014} proposed to add stimulus dependence to both restricted and semi-restricted Boltzmann machines (\cfeq\ref{eq:RBM}). In order to keep the number of stimulus-dependent parameters low, they considered a model where neuron the fields $\bm{a}(\bm{s})$ depend on the stimulus, but the couplings $\bm{w}$ with hidden units and hidden unit fields $\bm{b}$ do not. As above, a {time-}dependent $\bm{a}_t$ can be used instead of a \textit{stimulus-}dependent $\bm{a}(\bm{s})$ if stimulus repetitions are available.

\subsubsection{Joint stimulus-response ME model}
Another approach to capture the dependence between neurons and stimulus is to model the joint probability of stimuli $\bm{s}$ and responses $\bm{\s}$ with an ME model.
\citet{Gerwinn2009} for example proposed the \textbf{joint pairwise maximum entropy model}, an ME model with pairwise interactions between a continuous stimulus $\bm{s}$ and binary responses $\bm{\s}$. Similarly to the Ising and RBM models, this model has the form:
\begin{equation}
P(\bm{s},\bm{\s}) \propto\exp\!\left( \bm{h}_1 \bm{\s} + \bm{\s}^\intercal \bm{J}_1 \bm{\s} + \bm{h}_2 \bm{s} + \bm{s}^\intercal \bm{J}_2 \bm{s} + \bm{s}^\intercal \bm{J}_3 \bm{\s} \right)
\end{equation}
where vectors $\bm{h}_1$ and $\bm{h}_2$ control the mean of $\bm{\s}$ and $\bm{s}$, and matrices $\bm{J}_1$, $\bm{J}_2$ and $\bm{J}_3$ control the correlation between them.
They are all learned jointly from data. Conditioned by a response, the stimulus has a Gaussian distribution, and conditioned by a stimulus, responses have an Ising model distribution. The structure of the model puts hard constraints on the statistics of the stimulus. Once marginalized over responses, the stimulus distribution is a mixture of Gaussians, which might be problematic for example if the stimulus used in an experiment is not well approximated by one.

A variant is \textbf{partially dichotomized Gaussian model} \citep{Cox1999,Gerwinn2009}, which combines the ideas of the dichotomized Gaussian model and of the joint pairwise ME model. A latent variable $\bm{z}$ is considered, such that the joint distribution between the stimulus $s$ and the latent variable $\bm{z}$ is Gaussian. Then, neuron $i$ spikes if latent variable $z_i$ is positive. The distribution of stimuli marginalized over responses is also Gaussian, making it only suitable for specific experiment designs.

These models have convenient properties for analysis. For example one can easily compute the spike-triggered average of each neuron or group of neuron. However, it is hard to interpret directly the parameters of these models.

\subsubsection{Models of correlated noise}

We have presented models of stimulus-dependent neural populations where the noise could be relatively complex. But it has been suggested that the noise in some populations might have a simple form, typically a global modulation of firing rates \citep{Scholvinck2015cortical}. Some models have been proposed to capture this simple kind of noise. They are easier to use than previous models, as they tend to have simple structures.
In the following, we assume that neurons' firing rates are influenced by stimulus $\bm{s}$ through a deterministic sensory drive, $\bm{\phi}(\bm{s})$, characterized by the neurons' tuning curves.
Given a firing rate, neuron responses are noisy, as typically captured by a Bernoulli (for binary variable) or a Poisson (for spike counts) process. 
Here we assume that the firing rate $\bm{\lambda}$ is also a stochastic process, correlated across neurons. 
Such models are called \textbf{doubly stochastic Poisson}, or \textbf{Cox processes}. 

\citet{Arieli1996} suggested that the activity of cortical sensory neurons could be described by summing a stimulus drive to the spontaneous activity. \citet{Scholvinck2015cortical} showed that the variability of cat V1 neurons was mostly shared. Furthermore, an additive shared variability could partially account for the dependence of pairwise correlations on neural states.
The \textbf{rectified Gaussian model} (\citet{Banyai2016}, derived from \citet{Carandini2004}) captures this shared variability. Correlation between neurons is represented by a latent correlated variable $\bm{z}$ of dimension $N$, independent of the stimulus. The firing rate of neuron $i$ is:
\begin{equation}
\lambda_{i} = f_r \left( \phi_i(\bm{s}) + z_{i} \right)
\end{equation}
where the rate function  $f_r(x) =  k \left[x - V_0 \right]_+^\beta$ is a rectifying function. The latent variable $\bm{z}$ is a centered Gaussian variable with covariance learned from data. Importantly, $\bm{z}$ is independent from the stimulus. Yet, since the rate function $f_r$ is non-linear, the amplitude of the noise does depend on the stimulus. \citet{Banyai2016} applied this model to population of neurons in macaque V1, and found that it could account for part of the response variability.

Other works showed that the shared variability is not limited to additive interactions, but also takes the form of common multiplicative changes \citep{Ecker2014state}, possibly in the form of transient 50-100 ms packets of spiking activity \citep{Luczak2013gating}. In the \textbf{affine population modulation model} \citep{Lin2015}, the firing rates are correlated both in gain and in offset:
\begin{equation}
\lambda_{i} = z_{i}^{(1)} \phi_i(\bm{s})  + z_{i}^{(2)}
\end{equation}
where  $\bm{z}^{(1)}$ and $\bm{z}^{(2)}$  are latent variables with correlations across neurons. $\bm{z}^{(1)}$ has mean 1, whereas $\bm{z}^{(2)}$ has mean 0. In the extreme case where $\bm{z}^{(1)}$ is always 1 the noise is purely additive, whereas if $\bm{z}^{(2)}$ is always 0 the noise is purely multiplicative. If $\bm{z}^{(1)}$ is always 1 and $\bm{z}^{(2)}$ always 0, there is no shared variability. \citet{Lin2015} found that in cat V1, both additive and multiplicative common noise where involved, with relative importance varying in time.\\

\citet{Franke2016} considered a similar, although slightly  more complex model for correlations between cells:
\begin{equation}
\lambda_{i} = z_{i}^{(1)} \times f_i \left(  z_{i}^{(2)} \phi(s) + z_{i}^{(3)} \right).
\label{eq:multi_noise_souce}
\end{equation}
$ \bm{z}^{(1)}$, $ \bm{z}^{(2)}$ and $ \bm{z}^{(3)}$ are 3 vectors of correlated latent variables independent of the stimulus and of each other, with $\<\bm{z}^{(1)}\>=\bm{1}$, $\<\bm{z}^{(2)}\>=\bm{1}$, and  $\<\bm{z}^{(3)}\>=\bm{0}$. 
As in the rectified Gaussian model, when $f_i$ is nonlinear the amplitude of the noise due to $\bm{z}^{(2)}$ and $\bm{z}^{(3)}$ depends on the stimulus, which is not the case for $\bm{z}^{(1)}$. The parameters of the model can be fit using a multi-step procedure based on a series of mean squared difference minimizations. Interestingly, \citet{Franke2016} showed that for direction selective cells in the rabbit retina, the stimulus-dependent noise captured by $\bm{z}^{(2)}$ was beneficial to stimulus encoding.

At this point it should be emphasized that the latent variables $\bm{z}$ discussed in this section have a very different interpretation than for stimulus-independent model, where those variables accounted for stimulus-induced correlations. Here, latent variable only reflect noise correlations, even though their impact may depend on the stimulus itself because of nonlinearities.
Even for a fixed stimulus, latent variables can be used to account for two different sources of variability. The first one, $\bm{z}$, which we have discussed so far, is the variability across the population likely to occur in real life conditions, due to neural network dynamics, such that repetitions of a stimulus evoke different responses even if repetitions are close in time. The other source of variability, denoted by $\bm{\zeta}$, represents fluctuations in activity on the time scale of experiments (\eg fluctuations in neurons firing rate), which are likely due to experimental conditions.
Often, there is no clear distinction between $\bm{z}$ and $\bm{\zeta}$, although $\bm{\zeta}$ is constrained to have slow variations. In some models $\bm{z}$ and $\bm{\zeta}$ are summed to represent variability on short and long time scales respectively \citep{Park2015}.
The \textbf{modulated Poisson model} is an example where the firing rate is modulated by the latent variable $\bm{\zeta}$ \citep{Rabinowitz2015a}:
\begin{equation}\label{modulated}
\lambda_{i} =  \zeta_{i} \times \phi_i(\bm{s}) 
\end{equation}
This model is in principle equivalent to \cfeq\ref{eq:multi_noise_souce} with $\bm{z}^{(2)}=\bm{1}$ $\bm{z}^{(3)}=\bm{0}$, but with the additional constraint that $\bm{\zeta}_t$ varies slowly in time to reflect long-term fluctuations of the cell states. In practice, this condition is enforced by setting $\log \bm{\zeta}_t$ to be a Gaussian process with strong temporal correlations. \citet{Rabinowitz2015a} applied this model to  extracellular recordings of single auditory neurons in the primary auditory cortex and midbrain of anesthetized ferrets. They showed that the modulated Poisson model can be used to infer the influence of the stimulus on responses with a higher precision than a model without modulation.

 % 3_multibin
\section{Temporal correlations}
So far we have described  models of responses of $N$ neurons in a single time bin. Neural populations also have strong temporal correlations of their activity across different time bins, and a variety of models have been proposed to describe these correlation by spanning multiple time bins.

\subsection{Stimulus-independent models}
\subsubsection{Adding time to models of synchronous activity}
A natural way to generalize models for single time bins to responses spanning $B$ time bins is to consider that each neuron in each time bin as a different neuron. We obtain a population of $NB$ activity units, to which we can apply any model of activity proposed for a single time bin. This process is sometimes called \textbf{time spatialization} \citep{Buonomano2009}.  There are several examples of this strategy, such as the spatio-temporal Ising model \citep{Marre2009,Ganmor2011sparse,Koster2014}, the spatio-temporal reliable interaction model \citep{Ganmor2011sparse} or the spatio-temporal restricted Boltzmann machine and semi-restricted Boltzmann machine \citep{Koster2014}. This approach has a major drawback: it cannot be used to describe stationary activity (invariant to time translation). It cannot be used to generate longer spike trains than $B$, or to evaluate their likelihood.

Other models for spatio-temporal interactions are specifically designed for stationary distributions. \citet{Vasquez2012} presented the general form of \textbf{ME models with temporal interactions for stationary distributions}:
\begin{equation}
P(\{\bs_t\})\propto \exp\left[\sum_t \left(\bm{h}^\intercal \bs_t + \sum_{\tau=1}^u \bs_t^\intercal \bm{J}_\tau\bs_{t+\tau}\right)\right].
\end{equation}
The sum over time in the exponential ensures that this distribution is invariant to time translations, and is a general feature of ME models for stationary distributions. Constraining ME models to account for temporal stationarity reduces the number of parameters to learn compared to time spatialization. All time bins share the same field $\bm{h}$ and the coupling weight $\bm{J}$ only depend on the time delay $\tau$.
The model was applied to the salamander retina using brute-force gradient descent. It was shown that interactions over a 30 ms delay could predict well cross-correlations in the activity with a longer delay, up to 120 ms.
Although \citet{Nasser2013} described Monte Carlo methods to infer the parameters of these models, this inference procedure is computationally very intensive and has only been applied to small populations of neurons.

A simpler alternative is to focus on the evolution of the population count $K_t=K(\bs_t)$ instead of tracking individual neurons. \citet{Mora2015} proposed  a ME model reproducing the temporal correlations of the population count over a finite delay, $P(K_t, K_{t+\tau})$ for $\tau=1,..,u$:
\begin{equation}
P(\{\bm{\s}_t\}) \propto \exp\!\left( \sum_{t} \left[ h(K_t) +  \sum_{\tau=1}^{u} J_\tau (K_t, K_{t+\tau}) \right] \right),
\label{eq:PKt}
\end{equation}
where the function $\bm{h}$ controls the distribution of the population count, and the function $\bm{J}_\tau$ controls its temporal correlation at a delay $\tau$. Functions $\bm{h}$ and $\bm{J}$ are learned jointly from data. 
The model was used to show that for a population of 185 neurons in the rat retina, a memory span of only $u=4$ time bins corresponding to a 40 ms delay described well the dynamics of the population count, and in particular the durations and sizes of neural avalanches.
This model can be further simplified by considering the spike count as a continuous variable, and constrain only its mean and covariance. 
The corresponding ME model is then equivalent to a Gaussian process in the spike count $K_t$, which can be described by a Gaussian autoregressive process with memory $u$ \citep{Mora2015}. This simplification makes the model very easy to learn, but results in a poor description of the spike count distribution.

Restricted Boltzmann machines can also be adapted to stationary distributions. The \textbf{temporal restricted Boltzmann machine} \citep{Gardella2018}, is a model with hidden units in each time bin, witch can interact with neurons in different time bins:
\begin{equation}
 P(\{\bm{\s}_t,\bm{z}_t\}) \propto \exp\left[ \sum_t\left( \bm{a}^\intercal \bm{\s}_t + \bm{b}^\intercal\bm{z}_t + \sum_{\tau=1}^u\bm{z}_t^\intercal \bm{w}_\tau \bm{\s}_{t+\tau} \right)\right]
\end{equation}
It is a convolutional RBM, so interactions between neurons and hidden units only depend on the delay $\tau$ between them, not on their absolute time.
Importantly, once the model parameters are learned, they can be used to model responses of any time length. \citet{Gardella2018} found that this model was an accurate description of temporal correlations for a population of neurons in the rat retina. Furthermore, they showed that much information about the stimulus could be easily read out from the activity of hidden units.\\

Finally, it should be noted that the common-input models presented earlier for the synchronous activity can also be adapted to model stationary temporal correlations. \citet{Brette2009} presented spike train mixture methods to generate responses with fixed autocorrelations and cross-correlations between neurons, even in the case of non-binned spike trains. The dichotomized Gaussian model can also be generalized to account for temporal correlations \citep{Macke2009}.

\subsubsection{Latent dynamics}
In the temporal restricted Boltzmann machine, temporal correlations arise from interactions between neurons and hidden units with a time delay.
In a different approach, neural activities only depend on the current value of the latent variables, but these latent variables are correlated in time.
Temporal correlations between neurons then arise from temporal correlations in the dynamics of the latent variables. 
Models built on this idea are usually hard to learn, although expectation-maximization techniques have been proposed for this purpose \citep{Dempster1977,Yu2009}. This approach can be used to extract low-dimensional dynamics underlying neural responses, sometimes called neural trajectories \citep{Gao2016}. These trajectories are convenient for visualizing data and for decoding responses \citet{Yu2009}. They can also be used to decode movements from neural populations in motor areas \citet{Yu2009,Gao2016}.
As these models try to identify potentially simple latent dynamics underlying noisy responses, they are closely related to questions of dimensionality reduction (see \citet{Cunningham2014} for a review).
Much like the restricted Boltzmann machine, latent variables should be viewed as a convenient abstract tool rather than reflecting actual biological processes.
Latent variables can be continuous, typically with autoregressive Gaussian dynamics, or categorical, typically with Markov chain dynamics.

 In the \textbf{linear dynamical system}, the dynamics of the vector of latent variables $\bm{z}$ follows a Markovian Gaussian process:
\begin{equation}
\bm{z}_t = \bm{Az}_{t-1} + \bm{\epsilon}_t
\label{eq:autoGaus}
\end{equation}
where $\bm{A}$ is the dynamics matrix and $\bm{\epsilon}$ is the innovation noise, typically Gaussian. The neural trajectory $\bm{z}_t$ drives the firing rate of neuron $i$ through a rate function $f_i$:
\begin{equation}
\lambda_{it} = f_i(\bm{z}_t).
\label{eq:LDS}
\end{equation}
Stochastic responses are then produced by a Bernoulli, Poisson, or generalized count process. The dimensionality of the latent space is left to choice, and represents a trade-off between model accuracy and data required for learning. \citet{Gao2016} proposed a simple exponential nonlinearity $f_i$: 
\begin{equation}
\lambda_{it} = \exp \left(  \mu_i + \bm{C}^\intercal_i  \bm{z}_{t} \right)
\label{eq:LDSlin}
\end{equation}
where $\bm{C}_i$ is a projection vector and $\mu_i$ is a constant influencing the neuron's firing rate. Both $f+i$ and $\bm{C}_i$ are learned from data. More complex rate functions $f_i$ can be used for more accurate models. In the \textbf{linear dynamical system with nonlinear observation}, $f_i$ is adapted to each neuron and can be highly nonlinear. In \citet{Gao2016}, this is achieved by learning each $f_i$ with a feed-forward neural network model. \citet{Gao2016} shows that the resulting model is able to give a precise description of response dynamics in macaque V1. Furthermore, they are able to precisely correlate macaque hand-reaching directions to neural trajectories from the motor cortex.
\citet{Yu2009} used a similar model. They used the same firing rate vector $\bm{\lambda_t}$ driven by a latent variable (\cfeq\ref{eq:LDSlin}), but also allowed for interactions between neural responses in the same time bin. They did so by assuming that the square-root of the spike counts approximately followed a multivariate Gaussian distribution with mean $\bm{\lambda}_t$ and covariance $\bm{R}$ learned from data. The model was used to identify neural trajectories in macaque primary and premotor cortices during motion planning.

Unlike the latent dynamical system, \textbf{hidden Markov models} use latent variables with discrete states \citep{Abeles1995}, which are sometimes called ``modes.'' The stochastic dynamics of the latent variable $\bm{z}$ is typically characterized by a Markov chain, where the transition matrix is learned from data. The number of states can be high: for a population of 150 cells from a salamander retina, \citet{Prentice2016} reported that the model performance at describing data improved for up to 70 modes. 
For each latent state $\bm{z}$, responses are distributed according to an emission distribution $P(\bm{\s}|\bm{z})$.
Because one such distribution must be learned for each mode, the size of empirical data usually forbids emission distributions with many parameters.
For instance, \citet{Prentice2016} model emission distributions with independent neurons (with mode-dependent mean activities), or with Chow-Liu trees. A Chow-Liu tree is a simplification of the cascaded logistic model (\cfeq\ref{eq:CLM}), where the response of neuron $i$ only depends on just a single neuron $j<i$ previously characterized in the cascade. As the order of neuron indices is important for this model, it could be different for each latent mode. A Chow-Liu tree can only produce weak correlations for large populations of neurons, but it has a low number of parameters compared to the Ising and cascaded logistic models. 
In the recordings from the salamander retina, interactions added by the Chow-Liu tree structure resulted in improved predictions for correlations between pairs and triplets of neurons. The hidden Markov model resulted in a better description of responses than the reliable-interaction and $K$-pairwise models, as measured by the likelihood.
However, it is unclear how these models could generalize to larger populations, as all neurons from the population share the same latent mode, and the diversity of modes is expected to grow exponentially with the number of neurons, even when correlations are weak.

 % 4_transition
\subsection{Stimulus-dependent models}
\subsubsection{Transition models}

Neural networks are influenced by both stimulus and internal dynamics. We have presented models that account for each effect separately. We now present models where neurons are influenced by both. These models take the form of transition probabilities, which describe the response probability in a single time bin, given the population spiking history in previous time bins and the stimulus.
Neurons are often assumed to be independent given this history, and we can then model each neuron independently. We refer to such models as \textit{transition models}. The two kinds are the generalized linear model and the leaky integrate and fire neuron model.
All models presented here share a common structure: inputs drive a current $I$ entering the cell, and this current induces spikes. The input current is described by a linear sum of inputs. A stochastic process describes how the current induces spikes.
Models presented here should be considered as abstract descriptions of correlations between inputs and neural activities, rather than an actual causal model describing biological mechanisms driving neural responses. We first start by describing how the input current is influenced by the stimulus and the past activity of the population. 
Then we will present the different spike generation processes.

All correlations captured by the transition models originate from the inputs driving the current $I$. 
Here we describe the most common form,  where the input current of neuron $i$ in time bin $t$ is influenced additively by the stimulus and by the population activity in previous time bins $\bm{\s}_{t-\tau}$ for $\tau>0$ \citep{Pillow2005,Pillow2008}:
\begin{align}
 I_{it} = \phi_i(\bm{s}_t) +\sum_j \sum_{\tau=1}^{\tau_\text{max}}L_{ij,\tau} \s_{j(t-\tau)},
 \label{eq:GLM_I}
\end{align}
where $\phi_i(\bs)$ is the stimulus drive, and the filters $\bm{L}_{ij,tau}$ capture excitatory as well as inhibitory inputs from other cells with delay $\tau$. If the time bin is short enough, the filter $\bm{L}_{ii,\tau}$ can also be used to model the influence of the refractory period on the spiking activity.
The stimulus input is also learned from data. Usually, it is a linear spatial-temporal filtering of the stimulus \citep{Pillow2008}: $\phi_i( \bm{s}_t ) = \bm{k}_i^\intercal  \bm{s}_t$, where $k_i$ is a filter and $\bm{s}_t$ is a vector containing all the features of the stimulus at time $t$ as well as its recent past up to $t-\tau+1$. Quadratic functions of the stimulus have also shown to model accurately cat V1 cells \citep{Park2013spectral}.

The \textbf{generalized linear model} (GLM) is one of the most popular causal models, and is similar to the Linear Nonlinear Poisson model of neural response to stimuli \citep{Simoncelli2004}. In the GLM, the response of the cell is a stochastic process with a firing rate $\lambda_{i}$. This rate only depends on the current value of the input current $I$:
\begin{align}
\lambda_{it} = f_i\left(   I_{it} \right) 
\label{eq:GLM}
\end{align}
where $f_i$ is the rate function. Typically, the rate function is an exponential, as this choice drastically simplifies computations:
\begin{align}
f_i\left(   I \right) = \exp \left( \mu_i + I \right)
\end{align}
where $\mu_i$ is a parameter controlling the firing rate in the absence of input. $f_i$  can also be chosen to be a sigmoid function to account for saturation, or a function with a sub-exponential growth to avoid unrealistically high firing rates. Responses $\s_{it}$ are generated by a stochastic noise process with mean $\lambda_{it}$ (similarly to previously discussed models): Bernouilli for binary variables and Poisson or more general models of variability for spike counts \citep{Gao2015}.

Note that the firing rate at time $t$ depends on filtered population responses at time $t-1$. Thus, contrary to previous models with interactions between neurons in a same time bin, here there is no interaction between neurons in the same time bin. As a consequence, a GLM is simpler to learn than an Ising model, as it can be learned for each neuron independently.

The GLM is usually learned by maximizing the likelihood of observed responses. 
If the rate function $f_i$ is log concave, this problem is convex.
For small time bins where the firing rate is low, learning a GLM with a Bernoulli or Poisson process is equivalent \citep{Truccolo2005}. A number of priors can be added to the problem to constrain interactions to be sparse or filters to be smooth \citep{Stevenson2009}.
Using such priors, \citet{Stevenson2009} showed that groups of neurons in macaque primary motor and premotor cortices had similar inputs and impact on other neurons. These groups were interpreted as functional assemblies.	

As already discussed in the context of single-bin models of activity, It is possible to infer a \textit{time-}dependent model instead of a \textit{stimulus}-dependent model when a model to map stimulus to responses is not known, by replacing $\phi_i(\bm{s}_t)$ by a more general $\phi_{it}$ learned from data (\citet{Roudi2011a}, where the model is called kinetic Ising with time-dependent fields).
\citet{Cui2016inferring} showed that the GLM can be further improved by adding an input corresponding to a filter applied to the LFP. This is of particular interest, as \citet{Kayser2009} showed that taking the LFP phase into account drastically improved decoding from spike patterns.
	
GLMs can make precise predictions for responses in a single time bin, given the population response in previous time bins. This performance has been shown in multiple neural systems, such as macaque and human primary motor cortex \citep{Truccolo2005}, macaque retina \citep{Pillow2008} and macaque middle temporal area \citep{Cui2016inferring}. 

Although GLMs can predict single time bins, in general they cannot be used for generating data, as they tend to be very unstable \citep{Gerhard2017stability,Hocker2017multistep}. 
If one generates synthetic data by drawing the activity in each time bin conditioned on the synthetic data already generated in previous time bins, neural trajectories usually converge to fixed points with unrealistically high firing rates. 
\citet{Gerhard2017stability} proposed tools to analyze the stability of GLMs, along with inference methods to constrain GLMs to be stable. 
\citet{Hocker2017multistep} showed that stability could also be achieved by maximizing the likelihood of not only single time bins given their past, but also of the following time bins.

The \textbf{leaky integrate and fire model} is very similar to the GLM \citep{Bohte2000}, in the sense that it shares the same input current $I$ (\cfeq\ref{eq:GLM_I}). The difference lies in the spike generation process, which is closer to realistic biological mechanisms. The input current drives fluctuations in the membrane potential $V_{i}$, and a spike is emitted when the membrane potential reaches a threshold $V_\text{th}$. The membrane potential evolves according to:
\begin{equation}
V_{it} = (1 - \gamma) V_{i,t-1} + I_{it} + \epsilon_{it}
\label{eq:LIF}
\end{equation}
where $\gamma$ sets how fast the membrane potential goes back to 0 in the absence of input. $\epsilon$ is a random, usually Gaussian, fluctuation called noise innovation. When $V_{it}$ reaches $V_\text{th}$, $V_{i,t+1}$ is set back to 0. 
\citet{Cocco2009a} proposed efficient techniques to infer parameters by likelihood maximization. Using a simple parametrization of the spike-history filters, $L_{ij,\tau} = w_{ij} L_\tau$, they found that the interaction weights $w_{ij}$ correlated strongly with the coupling coefficients $J_{ij}$ inferred from an Ising model.
The same leaky integrate and fire model is used in the tempotron \citep{Gutig2006}, where it is trained to discriminate different correlations in input spike trains. 

\citet{Banyai2016} found that for cells in  macaque V1, the leaky integrate and fire model was a more accurate description of neural temporal correlations than the GLM. Multiple variations of the spiking process have been proposed \citep{Teeter2017}, with different degrees of similitude to biological processes, although more detailed models tend to be harder to learn.

In practice, the GLM is more commonly used in the literature than the leaky integrate and fire model. Although GLMs offer a coarser description of responses, and predict poorly some response statistics such as interspike intervals \citep{Hocker2017multistep}, they are more convenient to use and to learn from data.

\subsubsection{Transition models with latent dynamics}
The transition models presented above can account for some correlations between recorded neurons. Often, recorded neurons are part of a larger population, especially in the cortex. In this larger population, neural dynamics can have important effects. The effect of such dynamics can be included in more general transition models, in the form of dynamical latent variables. In practice, these models correspond to GLM rather than leaky integrate and fire neurons, as they are simpler to learn. 
This kind of model is sometimes called \textbf{generalized linear model augmented with a state-space model} \citep{Vidne2012}. 
\citet{Pillow2007} presented methods to infer such GLM in the simple case of a single recorded neuron interacting with a single latent neuron. Using simulations, they showed that taking into account the latent neuron improved model predictions.
Augmented GLM have also proven very successful at decoding and predicting correlations between neurons in the retina \citep{Vidne2012} in V1 \citep{Archer2014} or in motor cortex \citep{Lawhern2010,Macke2011}.

There are several ways to account for latent dynamics in the GLM input current (\cfeq\ref{eq:GLM_I}). A simple way is to take a latent variable $\bm{z}$ following a Gaussian autoregressive process (\cfeq\ref{eq:autoGaus}), and add the influence of the latent variable to the input current $I$ \citep{Kulkarni2007,Lawhern2010}.
\begin{equation}
I_{it} = \phi_i(\bm{s}_t) + \sum_j \sum_{\tau=1}^{\tau_\text{max}}L_{ij,\tau} \s_{j(t-\tau)}+ \bm{C}_i^\intercal \bm{z}_t 
\label{eq:GLMstimlatent}
\end{equation}
where $\bm{C}_i$ is a projection vector, measuring how the latent variable influences neuron $i$.
This model is a combination of the GLM (\cfeq\ref{eq:GLM_I}) with a linear dynamical system (\cfeq\ref{eq:LDSlin}). 
This model form is not easy to infer, but algorithms using expectation-maximization techniques have been proposed \citep{Dempster1977,Kulkarni2007}.

If the neural population is large, or if neurons are from different cortical areas, it can be interesting to use a latent space of high dimension. However, large latent spaces make the inference hard. 
To simplify the inference of latent dynamics, \citet{Vidne2012} constrained the latent dynamics to be composed of multiple independent dynamics of lower dimensions. In practice, this can be achieved by separating latent dimensions in batches, and constraining the transition coefficient $A_{nm}$ (governing the dynamics of latent variable, \cfeq\ref{eq:autoGaus}) to be 0 if $n$ and $m$ belong to different batches. \citet{Vidne2012} showed that for a population of 25 parasol cells in the macaque retina, the latent variable was more important than the population spiking history current $I_\text{pop}$ to reproduce neural correlations.

The stimulus is also likely to influence the dynamics of neighboring, unrecorded neurons. In some models, the stimulus also influences the latent variable dynamics:
\begin{equation}
\bm{z}_t = \bm{Az}_{t-1} + \bm{\varphi}(\bm{s}_t) + \bm{\epsilon}_t
\end{equation}
Here $\bm{\varphi}$ is a function of the stimulus, with same dimensionality as the latent variable.
For simplicity, the function $\bm{\varphi}$ is often chosen to be linear \citep{Archer2014,Park2015}, although more complex functions have been proposed, such as quadratic and with multiplicative interactions \citep{Archer2014}. 
Again, in the absence of a known model to relate stimuli to responses, a \textit{time-}dependent variable $\bm{\varphi}_t$ can be used instead of a \textit{stimulus-}dependent one \citep{Macke2011}.

In the above-mentioned form of the stimulus  (\cfeq\ref{eq:GLMstimlatent}), the direct stimulus input $\bm{\phi}$ can be included \citep{Macke2011}, or not \citep{Archer2014,Park2015}.
Models without direct stimulus input are convenient as they allow to have fewer receptive fields than there are neurons \citep{Archer2014}. This choice can significantly decrease the number of parameters to learn, and may be relevant for large populations of neurons in the cortex, where neurons share inputs from lower areas such as the retina or the cochlea. The population can then be considered as a ``computational unit'' \citep{Archer2014}.
	
Finally, as previously described for the modulated Poisson model (eq.~\ref{modulated}), it can be convenient to account for slow varying fluctuations that may be specific to experiments, using a latent variable $\bm{\zeta}$ constrained to have slow variations. This can be achieved by replacing $\bm{z}_t$ by $\bm{z}_t + \bm{\zeta}_t$ in \cfeq\ref{eq:GLMstimlatent}, while preserving the same dynamics for $\bm{z}$ \citep{Park2015}. \citet{Archer2014} showed that adding these slow variations improved the description of responses in  macaque V1 recordings.

 % 5_discussion
\section{Conclusion and future directions}

Over the last two decades, a variety of models for population responses has been proposed. 
Models can have very different forms, depending on analysis goals and experimental data available.
They can help us characterize and interpret features of the neural code, quantify the diversity of responses, separate the influence of the stimulus from internal effects, or estimate the robustness of stimulus encoding to response noise. 
Population models can provide an accurate description of the state of a neural population, even for single trials where single neurons might seem unreliable.
We have presented a large number and variants of population models, but this list is not exhaustive, and combinations of the aforementioned models are possible. 

The size of recorded populations increases at an unprecedented pace, and the neural system of some simple organisms can now be recorded entirely with increasing spatial and temporal resolution (see \citet{Prevedel2014} for \textit{C. elegans}, or \citet{Dupre2017} for the cnidarian \textit{Hydra vulgaris}). 
Enhanced inference techniques will be required to learn model on populations of thousands of neurons.
Future models will also have to adapt to future experimental constraints: the duration of recordings is not guaranteed to grow with the size of recorded populations; the sampling rate of individual neurons may be limited by the size of the recorded area when scanning techniques are used, such as in 2-photon microscopy. On the other hand, long recordings are not guaranteed to keep track of all neurons during the whole recording duration. Future models will need to account for neurons that are only observable in some parts of experiments. 

Models of neural dynamics will also need to adapt to larger populations. In the GLM for example, the number of couplings grows as the square of the population size, which will be hard to infer for large populations.
Fortunately, several works have suggested that the neural activity lies in a low dimensional manifold, and many dimensionality reduction techniques have been proposed to identify low-dimensional representations (see \citet{Cunningham2014} for a review).
Currently, GLMs model use complex dynamics for visible neurons, while other models use simple dynamics to describe latent variables of low dimensionality. Future works will need to model the dynamics of low-dimensional representations of observed neural responses.
Such models of neural dynamics will need to be translated into accurate models of spike trains. The framework of latent dynamical systems could be useful for this goal, using low-dimensional representations of the population activity instead of latent variables.

Most neural models presented here are applied to populations of neurons in a single area, for animals stimulated by a single stimulus.
Future model with need to account for interactions within each area along with interactions between multiple areas. 
A major challenge is also to account for the diversity of stimuli or even of neural states present in the recordings, and recapitulate this diversity in a single model. It is still challenging for a model to account for both awake animals and slow-wave sleep, as neurons behave differently in these regimes \cite{Nghiem2018}. 
Coarse-grained, modular models such as the hierarchical model \citep{Santos2010}, with different types of interactions at different spatial scales, offer promising avenues for this feat.
Another potential direction could be to develop models including both low-level recording techniques, such as electrode arrays and optical imaging, with high-level recordings, such as EEG and fMRI.

\bibliographystyle{natbib}

\end{document}